# On the muon transfer from protium to neon at low collision energies

S. V. Romanov[1,*]

[1]*National Research Centre "Kurchatov Institute", Moscow, 123182, Russia*

The direct muon transfer from protium to neon at low collision energies is considered. The question is discussed how the energy dependence of the transfer rate can be extracted from available experimental data. A model is suggested which is based on two assumptions. First, the muon transfer occurs within an interaction sphere out of which the entrance and transfer channels are not coupled. Secondly, for the $s$– and $p$–waves at low collision energies, the complex logarithmic derivatives of the radial wave functions of the entrance channel on the interaction sphere are energy-independent. With only the $s$–wave taken into account, two values of the logarithmic derivative were found, which agree with the transfer rates measured at the temperatures of 20 and 300 K. Within this temperature interval, the rates evaluated for these two values differ only slightly. Noticeable differences arise at temperatures of a few kelvins. The choice of the preferred value of the logarithmic derivative was made by the comparison of the energy dependence of the $S$–matrix with results of previous calculations. Thereby, the proposed procedure may be a way to correct calculations at low collision energies.

PACS numbers: 34.70.+e, 36.10.Ee

## 1. INTRODUCTION

The subject of the present study is the reaction of the direct muon transfer from protium to neon :

$$\mu p\,(1s) + \mathrm{Ne} \rightarrow \mu\mathrm{Ne}^* + p\,, \tag{1}$$

$\mu p\,(1s)$ is muonic protium in the $1s$–state, $\mu\mathrm{Ne}^*$ is muonic neon in an excited state. Let us introduce some quantities, which are usually used in considering such reactions. Let $\sigma_r$ be the total reaction cross section; it is summed over all the final states of muonic neon. Also, let $\sigma_{el}$ be the cross section of the elastic scattering of muonic protium by the neon atom. The above cross sections are functions of the collision energy $E$ or the relative velocity $v$ in the entrance channel: $v = \sqrt{2E/M}$, $M$ is the reduced mass. As a rule, the reaction rate $q_r$ is considered instead of the reaction cross

*Serguei.V.Romanov@gmail.com; Romanov_SVi@nrcki.ru

section:

$$q_r = N_{\rm H} v\, \sigma_r \,. \tag{2}$$

By tradition, it is reduced to the atomic density of liquid hydrogen $N_{\rm H} = 4.25 \times 10^{22}$ cm$^{-3}$. In experimental study of the muon transfer in dense gaseous mixtures, the rate $\lambda_r$ of the transfer from thermalized muonic atoms is usually measured. It is a function of the mixture temperature $T$. When reduced to the density $N_{\rm H}$ , it is related to the rate $q_r$ as follows:

$$\lambda_r = \int_0^\infty F(v,T)\, q_r\, dv \,, \tag{3}$$

$F(v,T)$ is the Maxwell velocity distribution function:

$$F(v,T) = \left(\frac{M}{2\pi k_B T}\right)^{3/2} 4\pi v^2 \exp(-Mv^2/2k_B T) \,. \tag{4}$$

It is normalized to unity:

$$\int_0^\infty F(v,T)\, dv = 1 \,, \tag{5}$$

$k_B$ is the Boltzmann constant.

Muonic protium is an electrically neutral object whose radius is two orders of magnitude less than radii of light atoms. In this respect, it is similar to the neutron. The direct muon transfer from protium to a heavier chemical element is an exoergic reaction. It is known [1] that for such reactions at low collision energies, when the $s$–wave predominates, the elastic cross section approaches a constant value, while the reaction cross section is inversely proportional to the relative velocity: $\sigma_r \propto 1/v$ . In this case, the reaction rate $q_r$ does not depend on the collision energy, and the rate $\lambda_r$ does not depend on temperature. The rate $\lambda_r$ of the reaction (1) was measured in liquid hydrogen-neon mixtures at temperatures of about 20 K [2] and in dense gaseous mixtures at room temperature (about 300 K) [3]. Results of these measurements given in Table 1 show that the reaction rate noticeably decreases with temperature. This means that the $1/v$ law is wrong at collision energies corresponding to room temperature. Moreover, the reaction (1) is strongly suppressed at room temperature: its rate is an order of magnitude less than the rates of the muon transfer to other chemical elements with the atomic number $Z \geq 6$ [2]. A qualitative explanation of this suppression was suggested in Ref. [4]. Within a modified version of the Landau—Zener model, it proved to be possible to attribute the decrease of the reaction rate to a destructive interference of waves in the reaction channels. Because of the above features, the reaction (1) is of special interest for theoretical study.

Table 1: Experimental $\lambda_{exp}$ and calculated $\lambda_{th}$ values of the rate $\lambda_r$ of the muon transfer from thermalized muonic atoms $\mu p\,(1s)$ to neon at two temperatures. All the rates are given in $10^{10}\,\mathrm{s}^{-1}$ and reduced to the atomic density of liquid hydrogen.

| $T$, K | $\lambda_{exp}$ | $\lambda_{th}$ | |
|---|---|---|---|
| | [2, 3] | [5] | [7] |
| 20 | $3.00 \pm 1.00$ | 2.00 | 1.2 |
| 300 | $0.849 \pm 0.018$ | 1.25 | 1.02 |

Results of two calculations of the rate $\lambda_r$ are also presented in Table 1. The first one was performed in Ref. [5] within a version of the perturbed stationary states method suggested in Ref. [6]. In particular, this method allows one to take into account the electron screening in the entrance channel by a simple modification of the potential of the polarization attraction between muonic protium and the neon nucleus. The values given in Table 1 were obtained for the case in which the electron screening was taken into account to a greater extent (the case C in Refs. [5, 6]). The second calculation was performed in Refs. [7, 8] within the method of hyperspherical elliptic coordinates. The electron screening was ignored in this calculation. The results of the second calculation agree better with the experimental value of the reaction rate obtained at room temperature. On the contrary, the first calculation better reproduces the value measured at the temperature of liquid hydrogen. Moreover, it more clearly demonstrates the decrease of the reaction rate with temperature. Of course, it should be borne in mind that the electon screening in a liquid mixture may be more complicated than assumed in Refs. [5, 6]. Moreover, the Maxwell distribution is a crude model in this case. Anyway, now we have the two experimental values of the reaction rate and the two calculations, which reproduce these values in a varying degree. [1] In this context, the question arises whether the dependence of the reaction rate on the collision energy in the region $E \lesssim 0.04$ eV can be extracted from available experimental data. Here the upper limit of $E$ is nearly equal to the mean thermal energy $(3/2)\,k_B T$ at room temperature. In order to solve this question, let us try to develop a model, which needs no extensive analysis of the reaction mechanism. This model will be based on results of the calculation performed in Ref. [5]. Therefore, let us consider

[1] A spread of theoretical results also exists for other elements with $Z \geq 6$. For example, the rates of the muon transfer to oxygen and nitrogen at room temperature were calculated in Refs. [7, 9]. For oxygen, the result of [7] agrees with experiment within 10 %, while the results of [9] differ from experimental values by a few times. The situation for nitrogen is opposite: the result of [9] agrees with experiment within 10 %, while the result of [7] exceeds experimental values by a factror of 1.5.

some of its relevant details. Unless otherwise specified, the muon-atom units are used below:

$$\hbar = e = m_\mu = 1\,, \tag{6}$$

where $e$ is the proton charge, and $m_\mu$ is the muon mass. The unit of length is $2.56 \times 10^{-11}$ cm, the unit of energy is 5.63 keV.

## 2. RESULTS OF PREVIOUS CALCULATIONS

The method used in Ref. [5] was based on a substantial difference of the energies of relative motion in the channels of the reaction (1). While the energy in the entrance channel $[\,\mu p\,(1s) + \mathrm{Ne}\,]$ is thermal, the energy gain in the transfer channel $[\,\mu\mathrm{Ne}^* + p\,]$ is a few keV [3]. It is clear that in this case an asymptotically correct description of the entrance channel should be provided first of all. Accordingly, the wave function of the three-body system (muon, proton, and neon nucleus) was constructed in the form of an expansion in eigenfunctions of a two-centre Coulomb problem, which was isolated in the three-body Hamiltonian written in the Jacobi coordinates of the entrance channel. These coordinates are the vector $\mathbf{r}$ specifying the position of the muon with respect to the proton, and the vector $\mathbf{R}$ connecting the centre of mass of muonic protium with the neon nucleus. The Coulomb centres are located at the ends of the vector $\mathbf{R}$; its length $R$ is the interatomic distance. The charge of one centre is equal to unity; it is at the centre of mass of muonic protium. The second centre is at the same point as the neon nucleus. Its charge is:

$$Z' = \frac{Z}{m} \approx 11.1\,, \tag{7}$$

$Z = 10$ is the neon charge, $m \approx 0.899$ is the reduced mass of muonic protium. In the limit of $R \to \infty$, the bound eigenstates of the two-centre problem fall into two groups. The states of the first group are localized near the centre carrying the unit charge. Their wave functions go into wave functions of isolated muonic protium with the correct reduced mass. The simplest approximation is to take into account the only state of the first group in the expansion of the three-body wave function. This state goes into the 1s–state of muonic protium at $R \to \infty$. This approximation already provides the correct dissociation limit in the entrance channel. Moreover, the polarization attraction between muonic protium and the neon nucleus is naturally obtained at large $R$, and the polarizability of muonic protium is reproduced with one percent accuracy. In the limit of $R \to \infty$, the states of the second group correspond to a muonic atom with an infinitely heavy nucleus carrying the charge $Z'$ rather than the real muonic atom of neon. It is clear that no partial cross sections of the muon transfer to individual states of muonic neon can be calculated in this case. Nevertheless, as at large $R$ the states of the second group are localized near the neon nucleus, their inclusion in the expansion of the three-body wave function allows one to describe the migration of the muon charge cloud from protium to neon and to calculate the total reaction cross section. Thus, the method used in Ref. [5] provides the asymptotically correct description of the entrance channel with low collision energies and removes flaws of the description into the transfer channel. The effect of these flaws is not expected to be too significant because of high energies of the relative motion in this channel.

In order to calculate the reaction and elastic cross sections within the approach considered, it is enough to find elements of the $S$–matrix in the entrance channel. The radial wave functions, which describe the relative motion in the reaction channels, satisfy a set of linear differential equations of the second order. As the eigenfunctions of the two-centre Coulomb problem are localized at different centres at $R \to \infty$ , the matrix elements of the interchannel interaction decrease exponentially at large $R$. As a result, the radial equations corresponding to the entrance channel are enclosed into a separate subset. In the simplest approximation, which was used in Ref. [5], this channel is described by the only two-centre eigenstate; it goes into the 1s–state of muonic protium at $R \to \infty$ . In this case, the entrance channel is asymptotically described by one equation:

$$\frac{d^2\chi_L}{dR^2} + 2M\,[\,E - U_{eff}(R)\,]\,\chi_L = 0\,, \tag{8}$$

$\chi_L(R)$ is the radial wave function, $L$ is the orbital angular momentum of the relative motion, $U_{eff}(R)$ is the effective potential energy:

$$U_{eff}(R) = U(R) + \frac{L(L+1)}{2MR^2}\,, \tag{9}$$

$U(R)$ is the potential energy of the interaction of muonic hydrogen with neon at large interatomic distances. The equation (8) is valid at $R \gtrsim 30$, where it is decoupled from the equations for the radial functions of the transfer channel. The boundary condition for the function $\chi_L$ at $kR \gg 1$ is:

$$\chi_L(R) \propto \exp\left[-\,i\left(kR - \frac{L\pi}{2}\right)\right] - S_L \exp\left[+\,i\left(kR - \frac{L\pi}{2}\right)\right]\,, \tag{10}$$

$k = \sqrt{2ME}$ is the momentum of the relative motion, $S_L$ is the diagonal $S$–matrix element. The reaction and elastic cross sections are:

$$\sigma_r = \sum_{L=0}^{\infty} \sigma_r^{(L)}\,, \qquad \sigma_{el} = \sum_{L=0}^{\infty} \sigma_{el}^{(L)}\,, \tag{11}$$

where the partial cross sections are expressed in terms of $S_L$ as follows [1]:

$$\sigma_r^{(L)} = \frac{\pi}{k^2}(2L+1)\left(1-|S_L|^2\right) , \tag{12}$$

$$\sigma_{el}^{(L)} = \frac{\pi}{k^2}(2L+1)\,|1-S_L|^2 \; . \tag{13}$$

The reaction rates $q_r$ and $\lambda_r$ can also be written in the form of partial expansions:

$$q_r = \sum_{L=0}^{\infty} q_r^{(L)} , \qquad q_r^{(L)} = N_{\mathrm{H}} v\, \sigma_r^{(L)} , \tag{14}$$

$$\lambda_r = \sum_{L=0}^{\infty} \lambda_r^{(L)} , \qquad \lambda_r^{(L)} = \int_0^{\infty} F(v,T)\, q_r^{(L)}\, dv\, . \tag{15}$$

If the electron screening is ignored, the interaction of muonic hydrogen with the neon nucleus at large $R$ is described by the potential of the polarization attraction:

$$U_p(R) = -\frac{\beta Z^2}{2R^4} , \tag{16}$$

$\beta = 9/2m^3 \approx 6.20$ is the dipolar polarizability of muonic hydrogen. With the electron screening taken into account, the potential energy $U(R)$ is:

$$U(R) = U_s(R) + U_f(R)\, . \tag{17}$$

The first term is the screened polarization potential:

$$U_s(R) = -\frac{\beta Z_a^2(R)}{2R^4} , \qquad Z_a(R) = Z - Z_e(R)\, , \tag{18}$$

$Z_e(R)$ is the absolute value of the electron charge in the sphere of radius $R$ with the centre at the neon nucleus, $Z_a(R)$ is the total charge of the neon atom in this sphere. The second term $U_f(R)$ is due to a finite size of muonic protium. It can be considered as a contact interaction of the muonic atom with the electron shell of neon:

$$U_f(R) = \frac{2\pi}{3} \langle\, r^2 \,\rangle\, \rho_e(R)\, , \tag{19}$$

$\langle\, r^2 \,\rangle$ is the mean square of the ground-state charge radius of muonic protium with respect to its centre of mass:

$$\langle\, r^2 \,\rangle = -\frac{3}{m}\left(1 - \frac{1}{M_{\mathrm{H}}}\right) , \tag{20}$$

$M_{\mathrm{H}}$ is the mass of the muon-atom nucleus (the proton mass for the reaction (1)). The function $\rho_e(R)$ is the absolute value of the electron density at the distance $R$ from the neon nucleus. It is normalized to the total nuclear charge:

$$4\pi \int_0^{\infty} \rho_e(R)\, R^2 dR = Z\, . \tag{21}$$

Table 2: The atomic charge $Z_a$ and the potential energies of the interaction of muonic hydrogen with neon versus the interatomic distance $R$. The values of $R$ and $Z_a$ are in muon-atom units, the potential energies are in eV. Numbers in parentheses are powers of 10. For example, $-2.30\,(-1) = -2.30 \times 10^{-1}$.

| $R$ | $Z_a$ | $U_p$ | $U_s$ | $U_f$ | $U$ |
|---|---|---|---|---|---|
| 25 | 9.13 | −4.46 | −3.72 | −2.30 (−1) | −3.95 |
| 30 | 8.86 | −2.15 | −1.70 | −1.46 (−1) | −1.85 |
| 40 | 8.47 | −6.81 (−1) | −4.88 (−1) | −6.33 (−2) | −5.51 (−1) |
| 50 | 8.15 | −2.79 (−1) | −1.85 (−1) | −3.16 (−2) | −2.17 (−1) |
| 75 | 7.50 | −5.51 (−2) | −3.10 (−2) | −1.31 (−2) | −4.41 (−2) |
| 100 | 6.75 | −1.74 (−2) | −7.94 (−3) | −9.42 (−3) | −1.74 (−2) |
| 125 | 5.83 | −7.14 (−3) | −2.43 (−3) | −6.80 (−3) | −9.23 (−3) |
| 150 | 4.87 | −3.44 (−3) | −8.18 (−4) | −4.67 (−3) | −5.49 (−3) |
| 175 | 3.97 | −1.86 (−3) | −2.93 (−4) | −3.11 (−3) | −3.40 (−3) |
| 200 | 3.17 | −1.09 (−3) | −1.10 (−4) | −2.04 (−3) | −2.15 (−3) |
| 250 | 1.95 | −4.46 (−4) | −1.70 (−5) | −8.73 (−4) | −8.90 (−4) |
| 300 | 1.16 | −2.15 (−4) | −2.92 (−6) | −3.78 (−4) | −3.81 (−4) |
| 400 | 0.397 | −6.81 (−5) | −1.07 (−7) | −7.52 (−5) | −7.53 (−5) |
| 500 | 0.133 | −2.79 (−5) | −4.91 (−9) | −1.62 (−5) | −1.62 (−5) |

The quantity $\langle r^2 \rangle$ is negative because it is mainly contributed by the negatively charged muon. Therefore, the potential $U_f(R)$ also corresponds to an attraction. The values of the potentials outlined above are presented in Table 2 for a number of interatomic distances. The electron density $\rho_e(R)$ and the charge $Z_e(R)$ were calculated with the help of analytical one-electron wavefunctions obtained within the Hartree-Fock-Roothan method [10]. The potential energies $U_s(R)$ and $U_f(R)$ decrease exponentially with $R$. As the function $U_s(R)$ is proportional to the square of the atomic charge $Z_a(R)$ and contains the additional factor $R^{-4}$, it decreases faster. As a result, this potential proves to be significant only at distances $R$ that do not exceed the electron Bohr radius (about 200 muon-atom units). For example, at $R = 30$ the potential $U_s = -1.7$ eV; it is an order of magnitude greater than $U_f$. As the electron $K$–shell of neon has nearly the same radius, the screening in $U_s$ is already noticeable: the atomic charge $Z_a(R)$ is about 8.9. At $R \approx 100$, the potentials $U_s$ and $U_f$ become equal, and their sum is about $-0.02$ eV. This value is of the order of thermal energies at room temperature. At $R = 200$, the potential $U_s$ is only 5 % of the potential $U_f$, which is equal to $-0.002$ eV. This value corresponds to thermal energies at the temperature of 20 K. One more feature of the potential $U_f$ should be noted. The electron screening makes the polarization attraction weaker, so that the inequality $|U_s| < |U_p|$ is valid at all values of $R$. The addition of the term $U_f$ leads to the potential $U$ becoming deeper than $U_p$ at $R > 100$: $|U| > |U_p|$. As $U$ decreases exponentially, while $U_p$ follows the power law $1/R^4$, the inequality sign is reversed at $R > 420$. In this point, the potentials are nearly equal to $-6 \times 10^{-5}$ eV that corresponds to a temperature of about 1 K.

Table 3: Parameters of curves of the effective potential energy for several values of the orbital angular momentum $L$. $R_0$ and $R_b$ are the positions of the zero and the potential-barrier top (in muon-atom units), $U_b$ is the barrier height in eV.

| $L$ | $R_0$ | $R_b$ | $U_b$ |
|---|---|---|---|
| 1 | 47.6 | 65.6 | 0.0672 |
| 2 | 28.9 | 40.5 | 0.577 |
| 3 | 21.0 | 29.5 | 2.17 |

Plots of the effective potential energy $U_{eff}(R)$ are shown in Figure 1. As the interatomic distance decreases, the potential curves with nonzero values of the orbital angular momentum $L$ pass through a maximum and then cross the zero-axis. The maximum acts as a barrier, which prevents the penetration of the respective partial wave into the region responsible for the muon transfer. The position $R_0$ of the zero of the function $U_{eff}(R)$, the position $R_b$ of the barrier top, and the barrier height $U_b$ are given in Table 3 for $L \leq 3$. For these values of $L$, the barrier top lies in the region of $R \gtrsim 30$, where the entrance channel is described by the equation (8). At $E \ll U_b$, the partial wave with the given value of $L$ makes a small contribution to the reaction cross section. As the collision energy grows, the partial cross section increases; at $E \sim U_b$ it becomes comparable to contributions of waves with lower orbital angular momenta. Moreover, quasi-steady states may exist in the subbarrier region. In a vicinity of such states, the reaction rate rises sharply.

The energy dependences of the reaction rate and the elastic cross section are plotted in Figures 2 and 3. The collision energy ranges between $10^{-5}$ and 15 eV. The lower boundary corresponds to a temperature of about 0.1 K, the upper boundary is close to the energy of the first electron exitation of the neon atom (16.6 eV [11]). At low collision energies, when the $s$–wave predominates, the reaction rate and the elastic cross section are nearly constant. As the collision energy grows, the contribution of the $s$–wave decreases but remains predominant up to the energy of $10^{-2}$ eV. At energies corresponding to room temperature, the $p$–wave becomes significant. At $E = 0.04$ eV, its contribution to the total reaction rate is about 25 %. The contribution of the $p$–wave to the elastic cross section is decisive because the partial cross section of the $s$–scattering passes through a minimum at $E \approx 0.05$ eV (the Ramsauer–Townsend effect). With further increase of the collision energy, the $d$–wave becomes decisive. In this wave, there is a quasi-steady state lying near the top of the barrier in the effective potential energy. As a result, a resonance peak arises on the curves at $E \approx 0.5$ eV. The effect of this peak is manifested up to the energy of 2 eV. At greater energies, partial waves with $L > 2$ become significant.

The temperature dependence of the rate of the muon transfer from thermalized muonic atoms is shown in Figure 4. At $T < 100$ K, the $s$–wave predominates, and the reaction rate decreases gradually. At greater temperatures, the $p$–wave becomes important; at $T = 300$ K it yields about 20 % of the total reaction rate. With further increase of temperature, the contribution of the $s$–wave still decreases, the contribution of the $p$–wave continues to increase, and, in addition, the $d$–wave becomes significant. As a result, the reaction rate passes through a wide minimum at $T = 300 - 700$ K and then begins to rise rapidly.

## 3. MODEL

In order to describe the muon transfer from protium to a heavier nucleus, let us take advantage of some views well known in nuclear physics [1]. Let us assume that the muon transfer occurs inside an interaction sphere of radius $R_c$, while at interatomic distances $R > R_c$ the entrance channel is described by one equation (8). The distance $R_c$ is the entrance channel radius. In accordance with those outlined in Section 2, let us suppose that $R_c = 30$ for the reaction (1). At a given value $L$ of the orbital angular momentum, the diagonal matrix element $S_L$ is determined by the complex logarithmic derivative (log–derivative) $f_L$ calculated on the surface of the interaction sphere:

$$f_L = R_c \, \frac{\chi_L'}{\chi_L} \, . \tag{22}$$

Here the radial function $\chi_L$ and its derivative $\chi_L'$ are taken at $R = R_c$; $f_L$ is a dimensionless quantity. The log–derivative is governed by conditions inside the interaction sphere, and a key point is its dependence on the collision energy. Let us confine ourselves to the consideration of the $s$– and $p$–waves, which make the main contribution to the reaction rate at thermal energies. For these waves, the entrance channel radius $R_c$ lies in a classically allowed region, and the absolute value of the effective potential energy in this point is 1–2 eV (Figure 1). Hence, the kinetic energy of the radial motion at $R = R_c$ is much greater than thermal energies. Therefore, it is natural to expect that at thermal energies the log–derivatives depend weakly on the collision energy and may be considered to be constant. This assumption is confirmed by results of the calculation discussed in Section 2. Figure 5 shows the energy dependences of the real and imaginary parts of the log–derivatives. Noticeable changes of these quantities arise only at $E \approx 0.1$ eV. At lower energies, the log–derivatives are nearly constant.

Let us accept the assumption that for the $s$– and $p$–waves at low collision energies the complex log–derivatives are energy-independent. Then their values should be fitted to the experimental values of the reaction rate (Table 1). As the log–derivatives are complex, there are four real parameters for the two partial waves, while only two experimental points are available. Therefore, let us adopt that the reaction rate is mainly contributed by the $s$–wave. This approximation is certainly valid at the temperature of liquid hydrogen. On the basis of results of the calculation [5], it may be expected that its accuracy at room temperature is 20–25 % (Section 2). As only the $s$–wave is considered below, let us omit the index $L = 0$ marking the quantities related to this wave. Then the reaction cross section is:

$$\sigma_r = \frac{\pi}{k^2}\left(1 - |S|^2\right) , \tag{23}$$

The $S$–matrix element can be written as follows:

$$S = |S| \exp(2\, i\, \delta) \, . \tag{24}$$

The phase $\delta$ varies within the interval $(-\pi/2\, ,\, \pi/2)$. Let us introduce some notations:

- $E_1 = 0.0025$ eV and $E_2 = 0.04$ eV are the mean thermal energies $(3/2)k_B T$ at the temperatures of $T_1 = 20$ K and $T_2 = 300$ K;
- $q_1$ and $q_2$ are the values of the reduced reaction rate $q_r$ at these energies;
- $S_1$ and $S_2$ are the values of the $S$–matrix element at these energies;
- $\lambda_1$ and $\lambda_2$ are the values of the rate $\lambda_r$ at the temperatures $T_1$ and $T_2$ .

The choice of the energy-independent log–derivative is made in three steps.

1. $q_1$ and $q_2$ are set equal to the values obtained in Ref. [5] for the $s$–wave. Then $q_1$ yields the absolute value of the element $S_1$ . Let $\delta_1$ be its phase. With the help of formulas given in Appendix A, the element $S_1$ is recalculated into the complex log–derivative, which yields the element $S_2$ and the reaction rate $q_r$ at the energy $E_2$ . At a given absolute value of $S_1$ , the rate $q_r$ is a function of the phase $\delta_1$ . The solutions of the equation

$$q_r(\delta_1) = q_2 \, , \tag{25}$$

provide the adopted values $q_1$ and $q_2$ of the reaction rate at the energies $E_1$ and $E_2$ . It was found that the equation had two solutions. Their values as well as the corresponding log–derivatives are listed in Table 4. Results obtained in Ref. [5] are also given there. They agree well with the first solution.

2. $q_1$ and $q_2$ are set equal to the experimental values of the reaction rate at the temperatures $T_1$ and $T_2$ (Table 1). The same procedure as at the step 1 yields two values of the phase $\delta_1$ again. One of them is positive, and another is negative. In order to establish a correspondence between these values and those obtained at the step 1 , let us require that the energy dependences of the real and imaginary parts of the $S$–matrix element be similar at each step. This can be achieved by grouping the solutions according to the sign of the phase $\delta_1$ as has been done in Table 4. The energy dependences of $\mathrm{Re}\,S$ and $\mathrm{Im}\,S$ obtained in this way are shown in Figures 6 and 7.

3. In order to refine the log–derivatives $f$ obtained at the step 2 , let us assume that the reaction rate $\lambda_r$ is a function of $\mathrm{Re}f$ and $\mathrm{Im}f$ , while the temperature is a parameter. Let us set the rates $\lambda_1$ and $\lambda_2$ equal to their experimental values (Table 1). Then $\mathrm{Re}f$ and $\mathrm{Im}f$ are solutions of the set of two equations:

$$\begin{aligned} \lambda_r(\mathrm{Re}f,\,\mathrm{Im}f,\,T_1) &= \lambda_1\,, \\ \lambda_r(\mathrm{Re}f,\,\mathrm{Im}f,\,T_2) &= \lambda_2\,. \end{aligned} \tag{26}$$

Two solutions of this set obtained by the Newton method are listed in Table 4. The log–derivatives found at the step 2 were used as initial approximations. This way provides a connection between the results obtained at the steps 2 and 3.

The two final values of the log–derivative found at the last step are linked to the results of Ref. [5]. The first solution seems to be preferable because it is in good agreement with these results at the step 1. It is interesting that for this solution the log–derivative changes considerably at each step; for the second solution the changes are noticeably less. Plots of the temperature dependence of the reaction rate $\lambda_r$ obtained for both the solutions are shown in Figure 8. In the range of 20–300 K , both the curves are close to each other, and the dependence of the reaction rate on the logarithm of temperature is nearly linear. Noticeable differences arise only at temperatures of a few kelvins. In this case, the values of the reaction rate exceed the values obtained in Ref. [5] by a factor of 1.5–2 , and the largest deviations occur for the first solution. Similar features occur for the energy dependence of the rate $q_r$ (Figure 9). In particular, there are also nearly linear parts of the curves in this case. Their slope corresponds to the dependence $q_r \propto 1/v$. This means that the reaction cross section is inversely proportional to the collision energy in accordance with general results obtained in Ref. [12].

Table 4: Resuts of searching for the energy-independent log–derivative $f$ for the $s$–wave. $\delta_1$ is the phase of the $S$–matrix element $S_1$ at the collision energy $E_1 = 0.0025$ eV. The results of Ref. [5] obtained for this energy are listed in the line marked with an asterisk. The final values of $\mathrm{Re}\, f$ and $\mathrm{Im}\, f$, which should be preferred, are taken in a frame.

| solution | step | $\delta_1$ | $\mathrm{Re}\, f$ | $\mathrm{Im}\, f$ |
|---|---|---|---|---|
| 1 | 1 | +0.193 | −14.4 | −2.25 |
| | * | +0.192 | −14.2 | −2.20 |
| | 2 | +0.370 | +10.1 | −0.641 |
| | 3 | +0.509 | +6.25 | −0.131 |
| 2 | 1 | −0.432 | +1.69 | −0.0161 |
| | 2 | −0.609 | +2.03 | −0.0175 |
| | 3 | −0.829 | +2.33 | −0.0121 |

Let us consider the elastic cross section. With only the $s$–wave taken into account, it is:

$$\sigma_{el} = \frac{\pi}{k^2} |1 - S|^2 = \frac{\pi}{k^2} \left(1 + |S|^2 - 2\,\mathrm{Re}\, S\right) ; \tag{27}$$

In contrast to the reaction cross section, which is specified only by the absolute value of $S$, the elastic cross section depends additionally on $\mathrm{Re}\, S$. For this reason, it proves to be more sensitive to the complex log-derivative. Plots of the energy dependence of $\sigma_{el}$ obtained for the two above solutions are shown in Figure 10. According to Ref. [5], the elastic cross section is mainly contributed by the $s$–wave at the collision energies less than $10^{-2}$ eV (Figure 3). In this range, the curves calculated for the two solutions are very different from each other and from the results of Ref. [5]. The curve obtained for the first solution goes below the curve corresponding to the second solution. Moreover, it lies noticeably above the curves from Ref. [5]. In the range of $E < 10^{-3}$ eV, where the curves tend to constant values, the elastic cross section calculated in Ref. [5] proves to be an order of magnitude less. Unfortunately, it is now impossible to use these resuts to separate the relevant solution because there are no experimental data on the elastic scattering of muonic protium by neon.

## 4. CONCLUSIONS

Let us briefly sum up the main points of consideration. The reaction of the direct muon transfer from protium to neon was considered at low collision energies, and an attempt was made to extract the energy dependence of the reaction rate from available experimental data. A model was suggested which is based on results of the previous calculation. It contains two assumptions:

1. The reaction occurs inside an interaction sphere. Outside this sphere, the reaction channels are uncoupled.

2. For the $s$– and $p$–waves at low collision energies, the complex log–derivatives of the radial functions of the entrance channel, which are calculated on the surface of the interaction sphere, are energy-independent.

The real and imaginary parts of the log–derivatives should be fitted to experimental values of the reaction rate. As there are only two experimental points obtained at the temperatures of 20 and 300 K, the fit was made with taking into account only the $s$–wave. The result proved to be ambiguous: two values of the log–derivative were found to agree with the experimental data. In the temperature range of 20–300 K, the reaction rates calculated for these values differ from each other by a few percent. Noticeable differences arise only at temperatures of a few kelvins. The choice of the relevant value of the log–derivative was made by comparing the energy dependences of the real and imaginary parts of the $S$–matrix element with the results of the previous calculation. Thus, as the experimental data are limited and the log–derivative is ambiguously determined, it is necessary to have some reference calculation. Then the proposed procedure may be a way of its correction at low collision energies. In this case, the log–derivative obtained in the calculation may be directly used as an initial approximation in the Newton method. This means that the step 2 may be omitted in the three-step sequence described in Section 3. However, the three-step sequence seems to be preferable because it provides a better initial approximation for the Newton method. Moreover, it is linked to the reference calculation in a less degree. This calculation is used only qualitatively — in order to separate the relevant solution by using the energy dependence of the $S$–matrix.

The consideration presented above was restricted to taking into account the $s$-wave contribution alone. As already mentioned, at room temperature the contribution of the $p$-wave may be 20–25 % of the total reaction rate. In order to separate this contribution, additional data are needed. For example, in Ref. [13] it was proposed to measure the rate of the muon transfer to oxygen at the temperatures of liquid nitrogen (77 K) and dry ice (195 K). According to the results of Ref. [5], at these temperatures the contribution of the $p$-wave to the rate of the muon transfer to neon is 6 % and 15 %, respectively. If experimental values of the rate of the muon transfer to neon were available at the temperatures of 20, 77, 195 and 300 K, it would be possible to consider both the $s$– and $p$–waves. In this case, the first two values allow one to determine the log–derivative for the $s$–wave, while the other two make it possible to extract the $p$–wave contribution. Moreover, an experimental study of the elastic scattering of muonic protium by neon is of interest because the elastic cross section is much more sensitive to the choice of the log–derivative than the reaction rate.

## Appendix A: LOG–DERIVATIVE AND $S$–MATRIX

Let us derive relations between the complex log–derivative and the $S$–matrix element in the entrance channel. The solution of the equation (8) at $R \geq R_c$ is written as follows:

$$\chi_L(R) = C_1\,\chi_1(R) + C_2\,\chi_2(R)\,. \tag{A1}$$

The index $L$ is omitted in the right side; $C_1$ and $C_2$ are complex constants, $\chi_1$ and $\chi_2$ are real fundamental solutions with the following asymptotic behaviour at $kR \gg 1$:

$$\begin{aligned}\chi_1(R) &\approx \sin\left(kR - L\pi/2\right),\\ \chi_2(R) &\approx \cos\left(kR - L\pi/2\right).\end{aligned} \tag{A2}$$

These solutions are constructed by a numerical integration of the equation (8) starting from the radius $R_c$. Their Wronskian does not depend on $R$; it is equal to $-k$. The transition to complex exponents yields at large $kR$:

$$\chi_L(R) \propto (-C_1 + i\,C_2)\exp\left[-i\left(kR - \frac{L\pi}{2}\right)\right] + (C_1 + i\,C_2)\exp\left[+i\left(kR - \frac{L\pi}{2}\right)\right]\,. \tag{A3}$$

The comparison of the formulas (10) and (A3) shows that

$$S_L = \frac{C_1 + i\,C_2}{C_1 - i\,C_2} = \frac{1 + iY}{1 - iY}\,, \tag{A4}$$

where $Y = C_2/C_1$ is a complex parameter. The log–derivative $f_L$ also includes $Y$:

$$f_L = R_c\,\frac{C_1\chi_1' + C_2\chi_2'}{C_1\chi_1 + C_2\chi_2} = R_c\,\frac{\chi_1' + Y\chi_2'}{\chi_1 + Y\chi_2}\,. \tag{A5}$$

Hereinafter the fundamental solutions and their derivatives are taken at the radius $R_c$. The relations between the log–derivative and the $S$–matrix element can be obtained by eliminating the parameter $Y$ from these formulas. For example, if the element $S_L$ is known, the formula (A4) yields:

$$\begin{aligned} \mathrm{Re}\, Y &= \frac{2\,\mathrm{Im}\, S_L}{(1+\mathrm{Re}\, S_L)^2 + (\mathrm{Im}\, S_L)^2}\,, \\ \mathrm{Im}\, Y &= \frac{1-|S_L|^2}{(1+\mathrm{Re}\, S_L)^2 + (\mathrm{Im}\, S_L)^2}\,. \end{aligned} \tag{A6}$$

Then the log–derivative is found from (A5):

$$\begin{aligned} \mathrm{Re}\, f_L &= \frac{R_c\left(\chi_1'\chi_1 + |Y|^2\,\chi_2'\chi_2 + \mathrm{Re}\, Y(\chi_1'\chi_2 + \chi_2'\chi_1)\right)}{(\chi_1+\chi_2\,\mathrm{Re}\, Y)^2 + (\chi_2\,\mathrm{Im}\, Y)^2}\,, \\ \mathrm{Im}\, f_L &= \frac{-kR_c\,\mathrm{Im}\, Y}{(\chi_1+\chi_2\,\mathrm{Re}\, Y)^2 + (\chi_2\,\mathrm{Im}\, Y)^2}\,. \end{aligned} \tag{A7}$$

If the log–derivative is known, the parameter $Y$ is expressed from (A5):

$$\begin{aligned} \mathrm{Re}\, Y &= \frac{R_c\,\mathrm{Re}\, f_L\,(\chi_1'\chi_2 + \chi_2'\chi_1) - |f_L|^2\,\chi_1\chi_2 - R_c^2\,\chi_1'\chi_2'}{(\chi_2\,\mathrm{Re}\, f_L - R_c\,\chi_2')^2 + (\chi_2\,\mathrm{Im}\, f_L)^2}\,, \\ \mathrm{Im}\, Y &= \frac{-kR_c\,\mathrm{Im}\, f_L}{(\chi_2\,\mathrm{Re}\, f_L - R_c\,\chi_2')^2 + (\chi_2\,\mathrm{Im}\, f_L)^2}\,. \end{aligned} \tag{A8}$$

The $S$–matrix element is found from (A4):

$$\begin{aligned} \mathrm{Re}\, S_L &= \frac{1-|Y|^2}{(1+\mathrm{Im}\, Y)^2 + (\mathrm{Re}\, Y)^2}\,, \\ \mathrm{Im}\, S_L &= \frac{2\,\mathrm{Re}\, Y}{(1+\mathrm{Im}\, Y)^2 + (\mathrm{Re}\, Y)^2}\,. \end{aligned} \tag{A9}$$

Let us note some features of the quantities introduced above. As $|S_L| < 1$ in the presence of a reaction, the inequality $\mathrm{Im}\, Y > 0$ follows from (A6). From (A7) it is obvious that $\mathrm{Im}\, Y$ and $\mathrm{Im}\, f_L$ are opposite in sign, i.e. $\mathrm{Im}\, f_L < 0$.

---

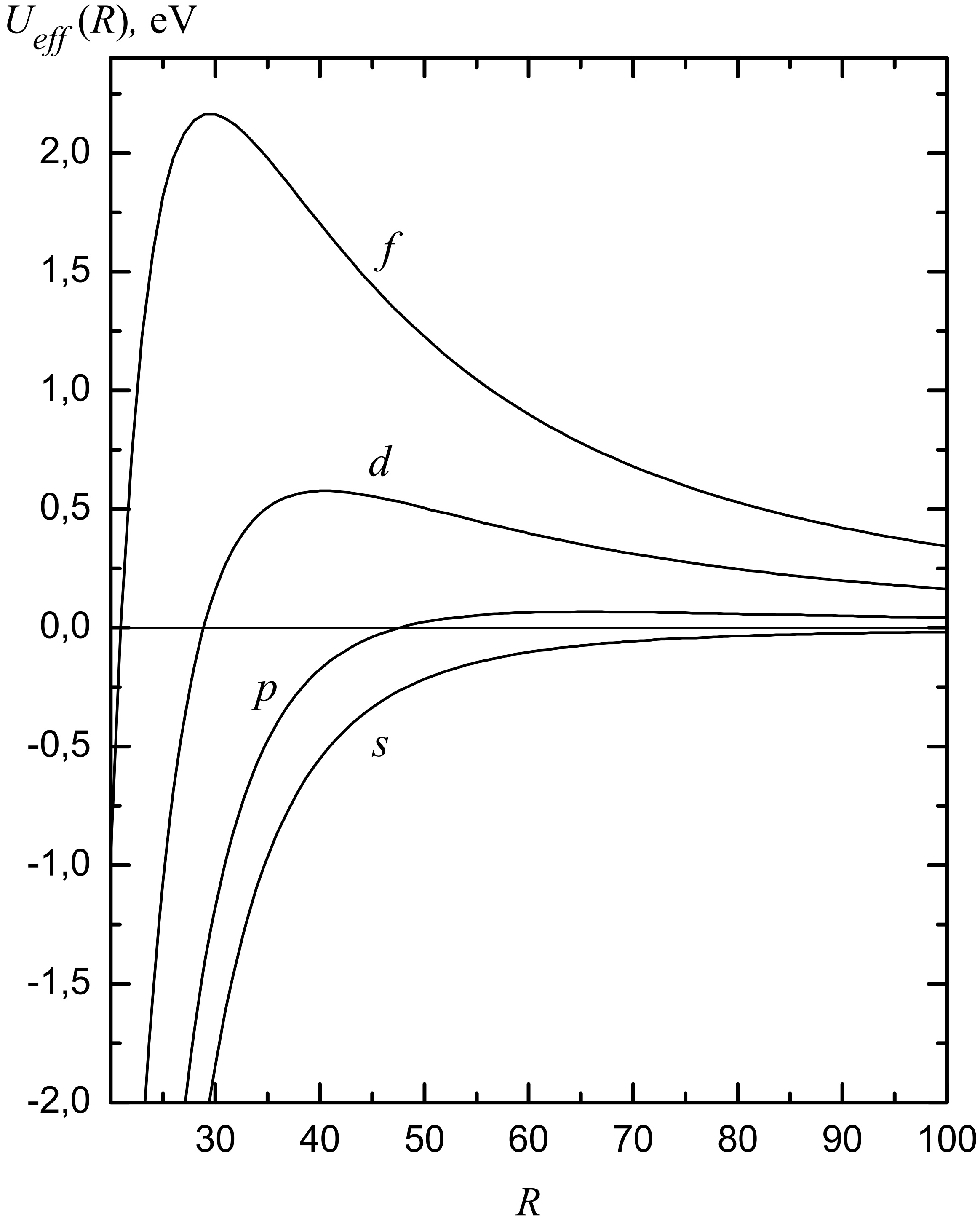

Figure 1: The effective potential energy $U_{eff}(R)$ for the $s$, $p$, $d$, and $f$–waves versus the interatomic distance $R$. The distance $R$ is given in muon-atom units. The $s$–wave curve represents the potential energy $U(R)$.

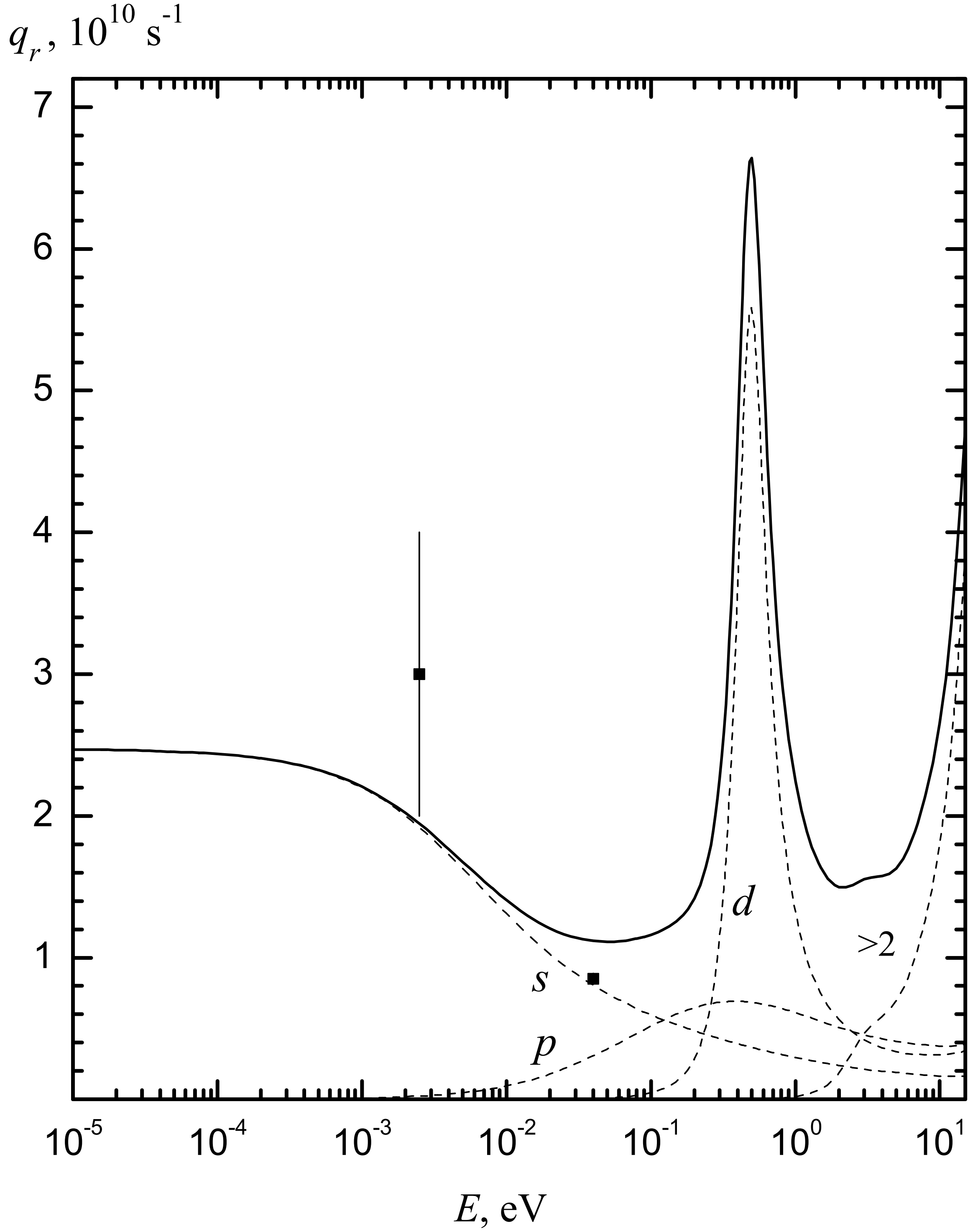

Figure 2: Reduced reaction rates versus the collision energy [5]. The solid curve is the total rate $q_r$ , the dashed curves are the partial rates $q_r^{(L)}$ for the $s$, $p$, and $d$–waves. The dashed curve marked as ' $> 2$ ' is the contribution of the waves with the orbital angular momenta $L > 2$ . The experimental values from Table 1 (black squares) are associated with the mean thermal energies $(3/2)\, k_B T$ at the temperatures of 20 and 300 K .

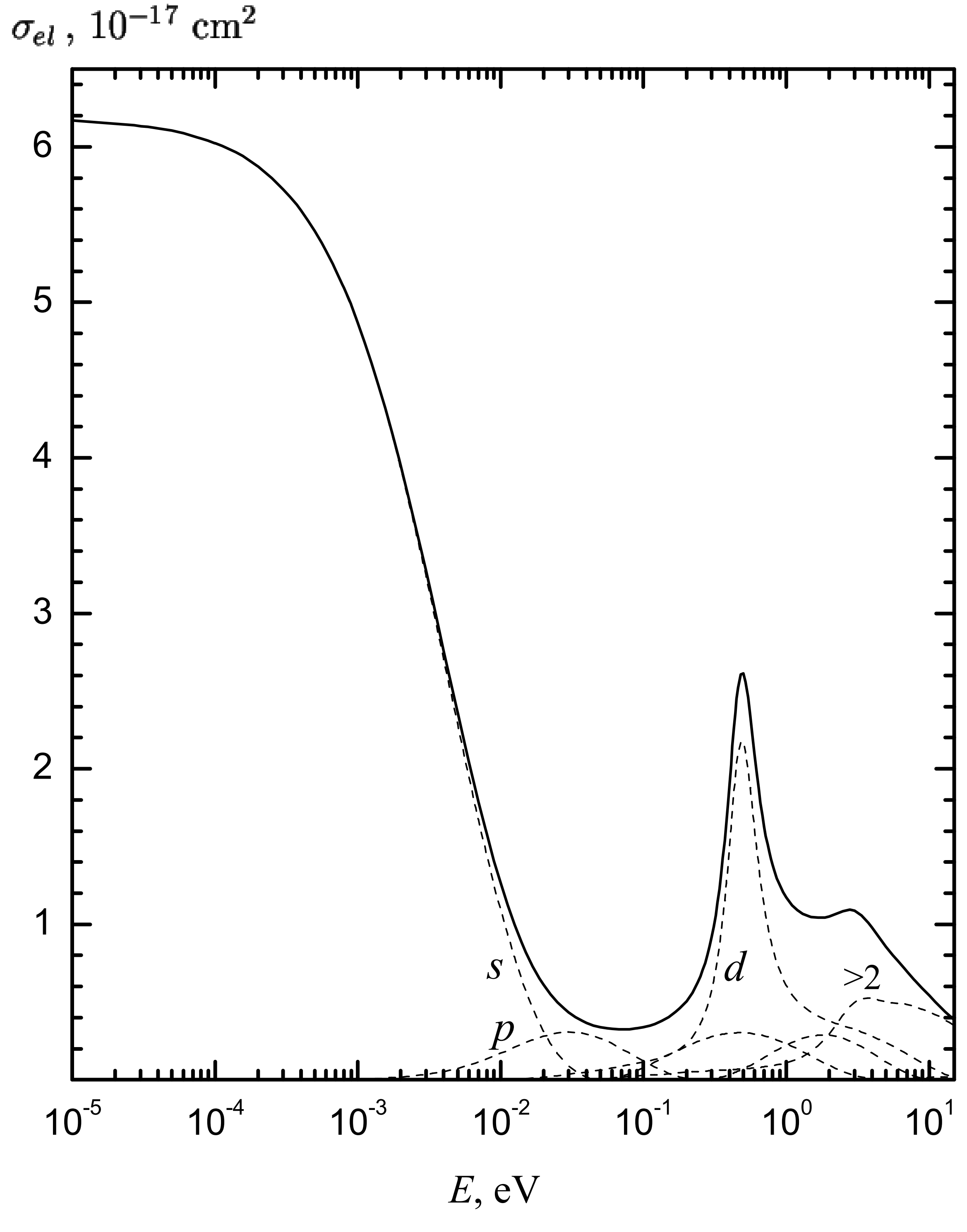

Figure 3: Cross sections of the elastic scattering versus the collision energy [5]. The solid curve is the total cross section $\sigma_{el}$ , the dashed curves are the partial cross sections $\sigma_{el}^{(L)}$ for the $s$, $p$, and $d$–waves. The dashed curve marked as ' $> 2$ ' is the contribution of the waves with the orbital angular momenta $L > 2$ .

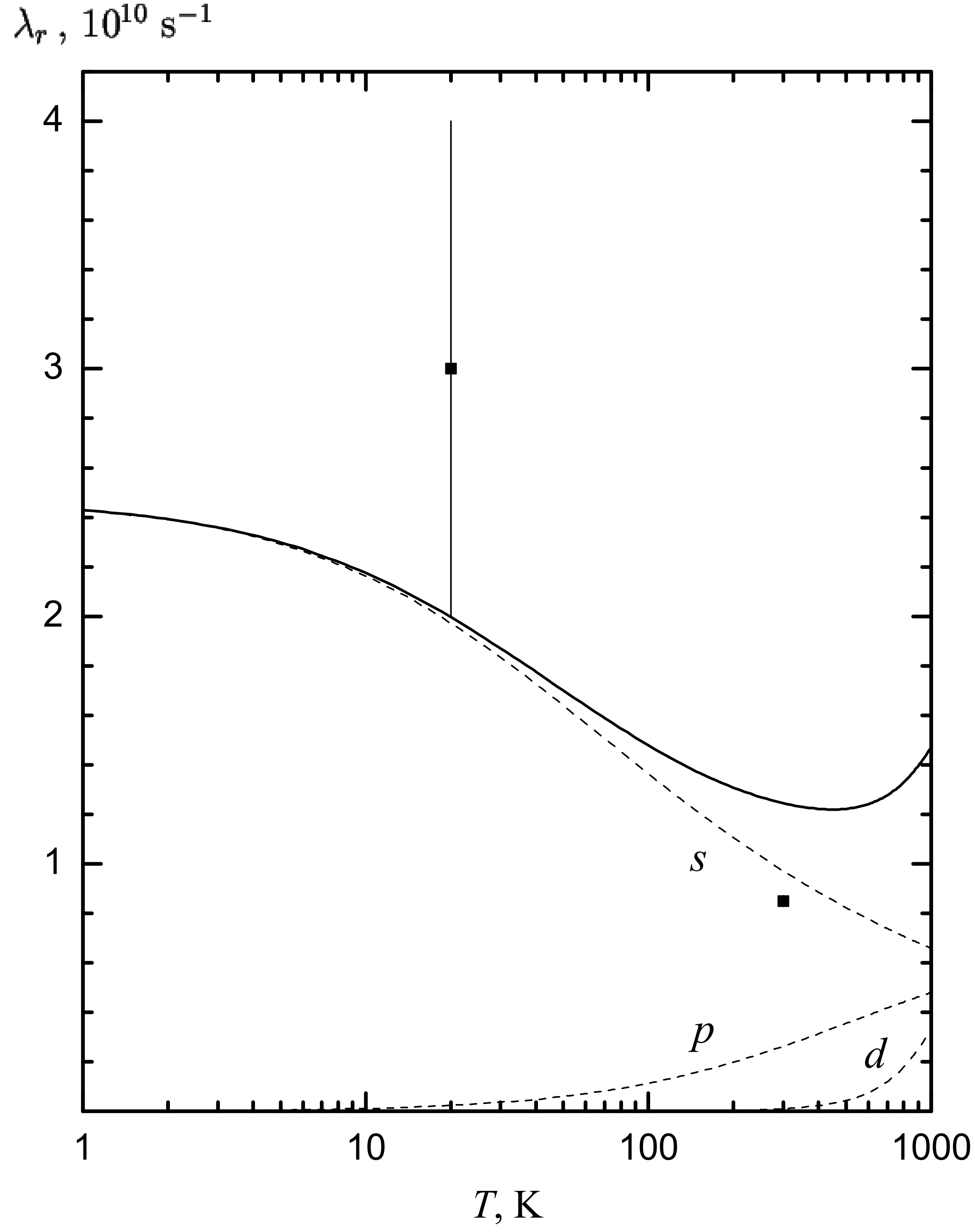

Figure 4: Reduced rates of the muon transfer from thermalized muonic atoms versus temperature [5]. The solid curve is the total rate $\lambda_r$ , the dashed curves are the partial rates $\lambda_r^{(L)}$ for the $s$, $p$, and $d$–waves. The experimental values (black squares) correspond to the temperatures of 20 and 300 K (Table 1).

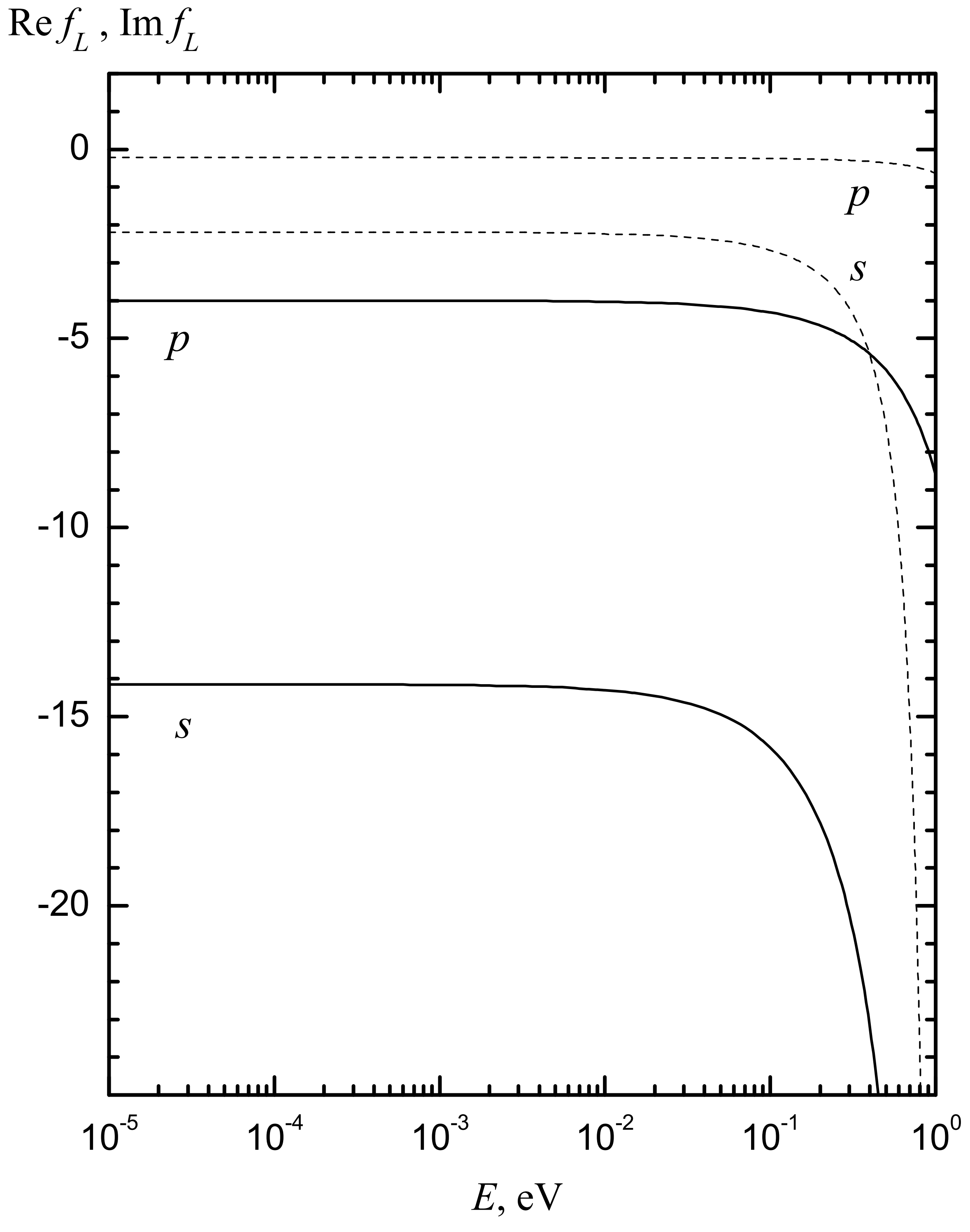

Figure 5: The real and imaginary parts of the complex log–derivatives for the $s$– and $p$–waves versus the collision energy [5]. The solid curves are the real parts, the dashed curves are the imaginary parts.

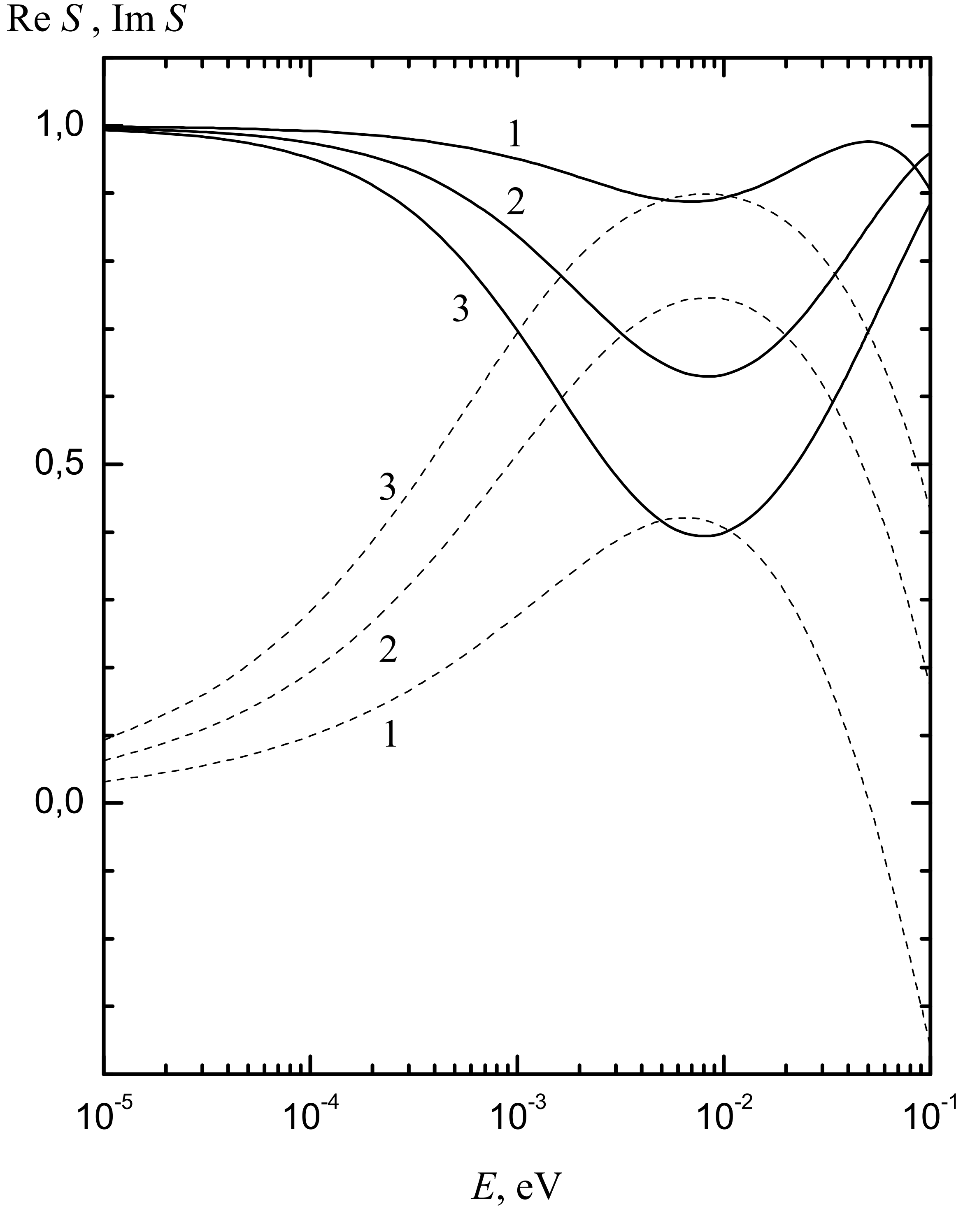

Figure 6: Real and imaginary parts of the $S$–matrix element versus the collision energy (the $s$–wave); the first solution from Table 4. The solid curves are the real parts, the dashed curves are the imaginary parts. Each curve is marked with the number of the step at which it was obtained.

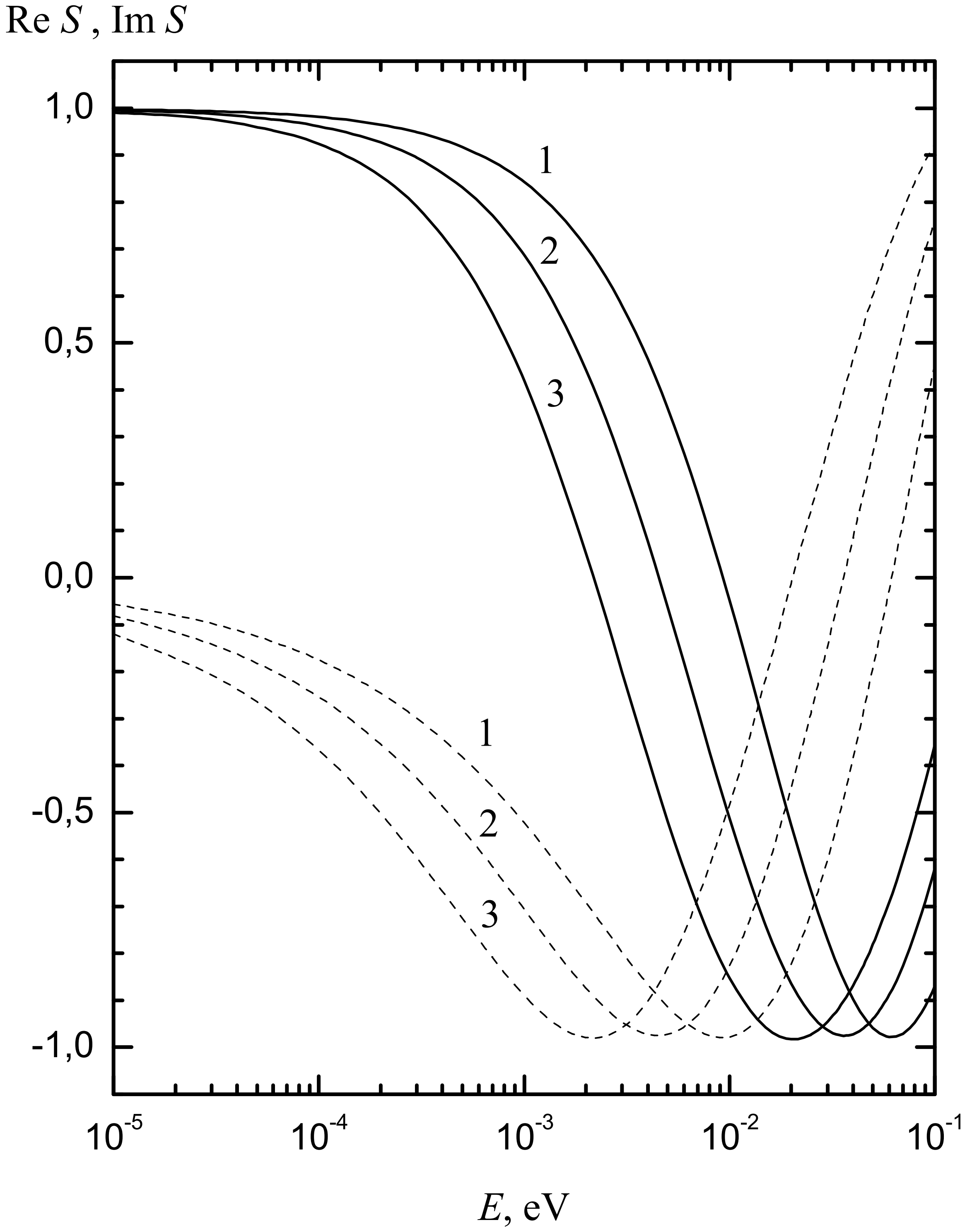

Figure 7: Real and imaginary parts of the $S$–matrix element versus the collision energy (the $s$–wave); the second solution from Table 4. The notations are the same as in Figure 6.

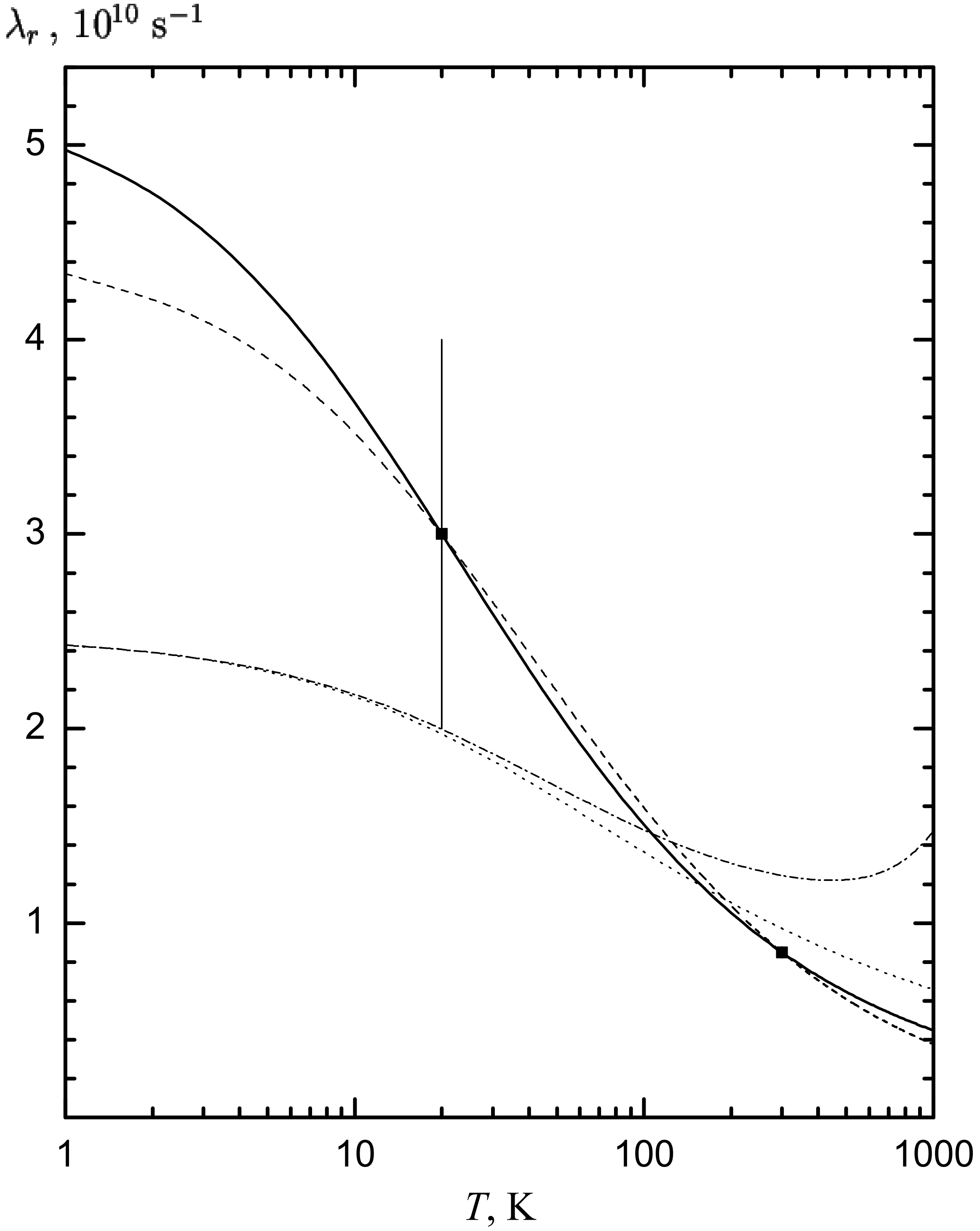

Figure 8: Reduced rates of the muon transfer from thermalized muonic atoms versus temperature. The solid curve is the first solution from Table 4, the dashed curve is the second one. The dash-dot and dotted curves are respectively the total rate $\lambda_r$ and the partial contribution $\lambda_r^{(L=0)}$ of the $s$–wave obtained in Ref. [5] (Figure 4). The experimental values (black squares) correspond to the temperatures of 20 and 300 K.

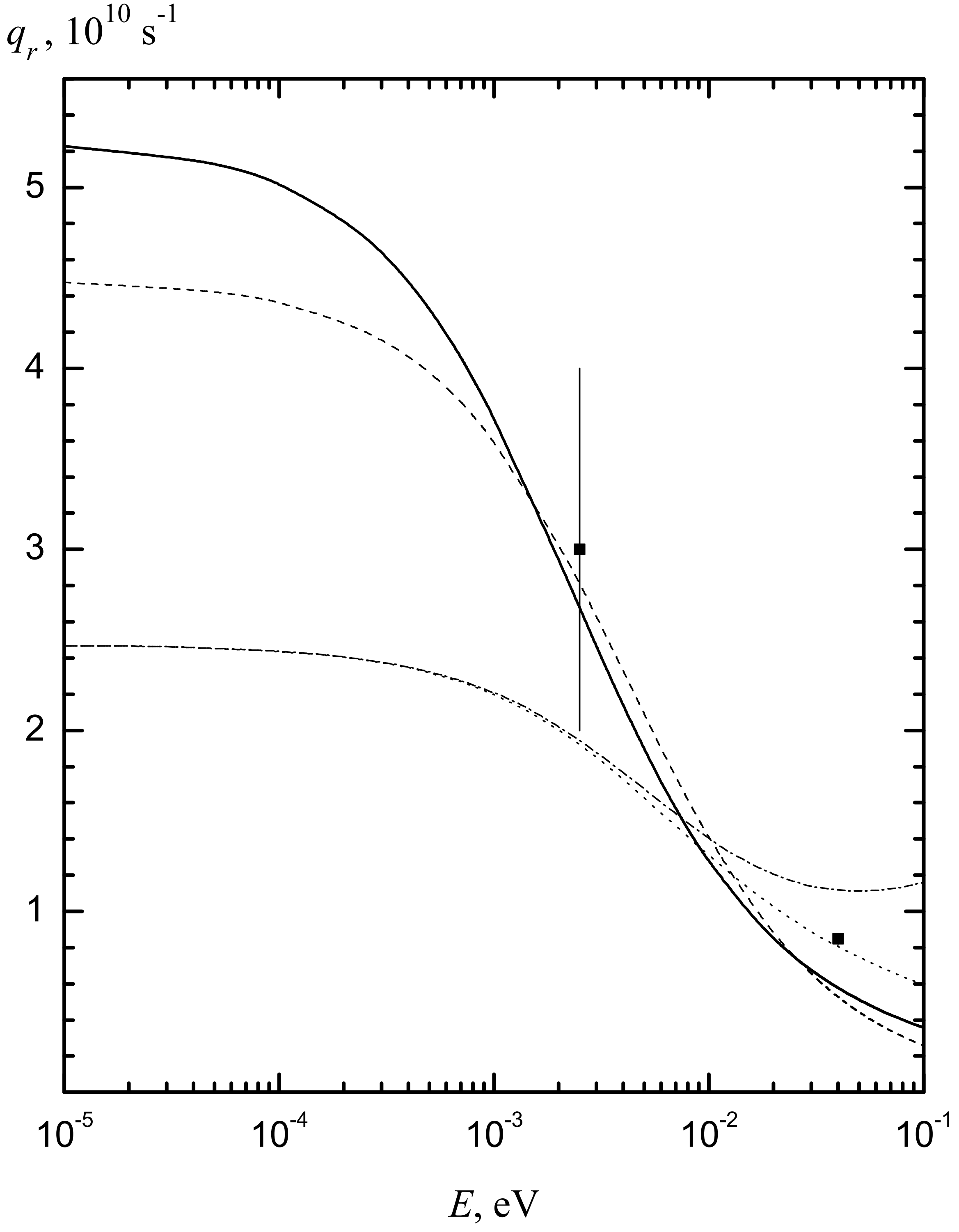

Figure 9: Reduced reaction rates versus the collision energy. The solid curve is the first solution from Table 4, the dashed curve is the second one. The dash-dot and dotted curves are respectively the total rate $q_r$ and the partial contribution $q_r^{(L=0)}$ of the $s$–wave obtained in Ref. [5] (Figure 2). The experimental values (black squares) are associated with the mean thermal energies $(3/2)\, k_B T$ at the temperatures of 20 and 300 K.

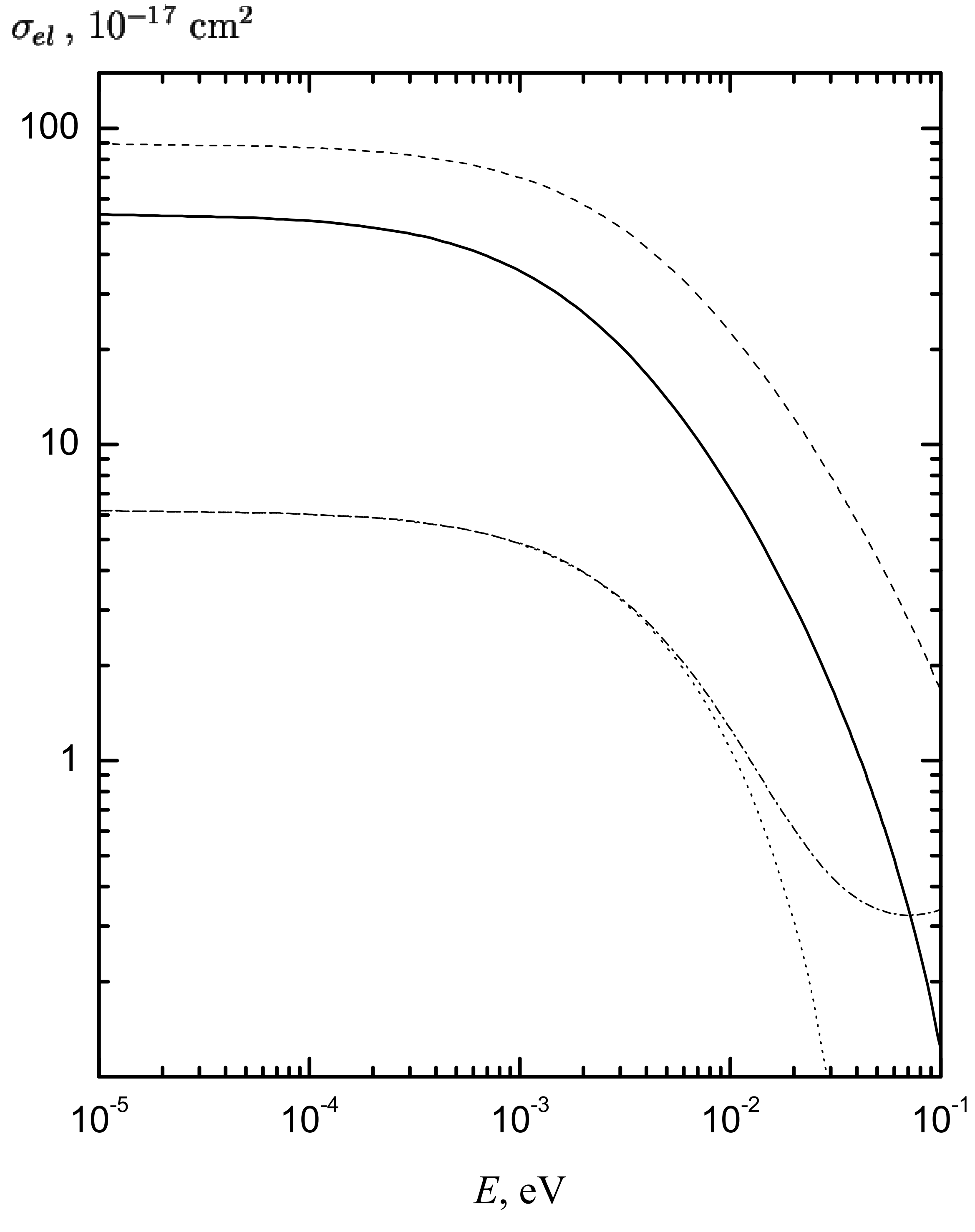

Figure 10: Cross sections of the elastic scattering versus the collision energy. The solid curve is the first solution from Table 4, the dashed curve is the second one. The dash-dot and dotted curves are respectively the total elastic cross section $\sigma_{el}$ and the partial contribution $\sigma_{el}^{(L=0)}$ of the $s$–wave obtained in Ref. [5] (Figure 3).